\documentclass[preprint,showpacs,titlepage,aps,prd,tightenlines
,amsmath,byrevtex,nofootinbib]{revtex4}

\usepackage{graphicx}

\usepackage[pdftex,urlcolor=blue]{hyperref}

\usepackage[normalem]{ulem}

\def\lsim{\raise0.3ex\hbox{$\;<$\kern-0.75em\raise-1.1ex
\hbox{$\sim\;$}}}
\def\gsim{\raise0.3ex\hbox{$\;>$\kern-0.75em\raise-1.1ex
\hbox{$\sim\;$}}}

\keywords{Neutrino Physics, Reactor Experiments}

\begin{document}

\vspace*{-1cm}
\title{ What is $\Delta m^2_{ee}$ ?}

\author{Stephen~Parke}
\email{parke@fnal.gov} 
\affiliation{Theoretical Physics Department, Fermi National Accelerator Laboratory, P.\ O.\ Box 500, Batavia, IL 60510, USA \\ }
\preprint{FERMILAB-PUB-16-020-T}

\vglue 1.6cm

\begin{abstract}
The current short baseline reactor experiments, Daya Bay and RENO (Double Chooz) have measured (or are capable of measuring) an effective $\Delta m^2$ associated with the atmospheric oscillation scale of 0.5 km/MeV in electron anti-neutrino disappearance. In this paper, I compare and contrast the different definitions of such an effective  $\Delta m^2$ and argue that the simple, L/E independent, definition given  by  $\Delta m^2_{ee} \equiv  \cos^2 \theta_{12} \Delta m^2_{31}+  \sin^2 \theta_{12} \Delta m^2_{32}$, i.e. ``the $\nu_e$ weighted average of $\Delta m^2_{31}$ and $\Delta m^2_{32}$,'' is superior to all other definitions and is useful for both short baseline experiments mentioned above and for the future medium baseline experiments JUNO and RENO 50.
\end{abstract}

\date{January 26, 2016}


\maketitle

\section{Introduction}
The short baseline reactor experiments, 
Daya Bay \cite{An:2012eh},
RENO \cite{Ahn:2012nd},
 and Double Chooz  \cite{Abe:2012tg} , 
 have been very successful in determining the 
electron neutrino flavor content of the neutrino mass eigenstate with the smallest amount of $\nu_e$, the state usually labelled $\nu_3$. The parameter which controls the size of this flavor content is the mixing angle $\theta_{13}$, in the standard PDG convention\footnote{A more informative notation for mixing angles $(\theta_{12}, ~\theta_{13}, ~\theta_{23})$ is $(\theta_{e2}, ~\theta_{e3}, ~\theta_{\mu 3})$, respectively, such that $U_{e2} = \cos\theta_{e3} \sin \theta_{e2}$, $U_{e3} = \sin \theta_{e3} e^{-i\delta}$ and $U_{\mu3} = \cos\theta_{e3} \sin \theta_{\mu 3}$.}, and the current measurements indicate that $\sin^2 2 \theta_{13} \approx 0.09$ with good precision ($\sim5$\%).

The mass of the $\nu_3$ eigenstate, has a mass squared splitting from the other two mass eigenstates, $\nu_1$ and $\nu_2$, of approximately $\pm2.4 \times 10^{-3} ~{\rm eV}^2$ given by  $\Delta m^2_{31} \equiv m^2_3 -m^2_1$ and $\Delta m^2_{32} \equiv m^2_3 -m^2_2$, the sign determines the atmospheric mass ordering. The mass squared difference between, $\nu_2$ and $\nu_1$, $\Delta m^2_{21} \equiv m^2_2-m^2_1 \approx +7.5 \times 10^{-5} ~{\rm eV}^2$ is about 30 times smaller than both $\Delta m^2_{31}$ and $\Delta m^2_{32}$, hence  $\Delta m^2_{31} \approx\Delta m^2_{32}$.
However, the difference between $\Delta m^2_{31}$ and $\Delta m^2_{32}$ is $\sim$3\%.

Recently, two of these reactor experiments, Daya Bay, see \cite{An:2013zwz} - \cite{An:2015rpev2} and RENO \cite{RENO:2015ksa}, have extended their analysis of their data, from just fitting $\sin^2 2\theta_{13}$, to a two parameter fit of both $\sin^2 2\theta_{13}$ and an effective $\Delta m^2$. The measurement uncertainty on this effective $\Delta m^2$ is approaching the difference between $\Delta m^2_{31}$ and $\Delta m^2_{32}$. So it is now a pertinent question ``What is the physical meaning of this effective $\Delta m^2$?''  Clearly, the effective  $\Delta m^2$ measured by these experiments is some combination of $\Delta m^2_{31}$ and $\Delta m^2_{32}$. Answering the  question  ``What is the combination of $\Delta m^2_{31}$ and $\Delta m^2_{32}$ is measured in such a short baseline reactor experiment?'' is the  primary purpose of this paper,

The outline of this paper is as follows: in Section \ref{sec:npz} the $\bar{\nu}_e$ survival probability is calculated in terms of an effective $\Delta m^2$ which naturally arises in this calculation, then this definition is applied to the short baseline reactor experiments, $L/E <$ 1 km/GeV.  In Section \ref{sec:otherdmsqs}, I review other possible definitions of an effective $\Delta m^2$, including the two invented by the Daya Bay collaboration. These  new effective $\Delta m^2$'s are either essentially equal to the effective $\Delta m^2$ of section \ref{sec:npz} or are $L/E$ dependent. This is followed by a conclusion and two appendices.


\begin{figure}
\begin{center}
\vspace{-1cm}
\includegraphics[width=0.6\textwidth]{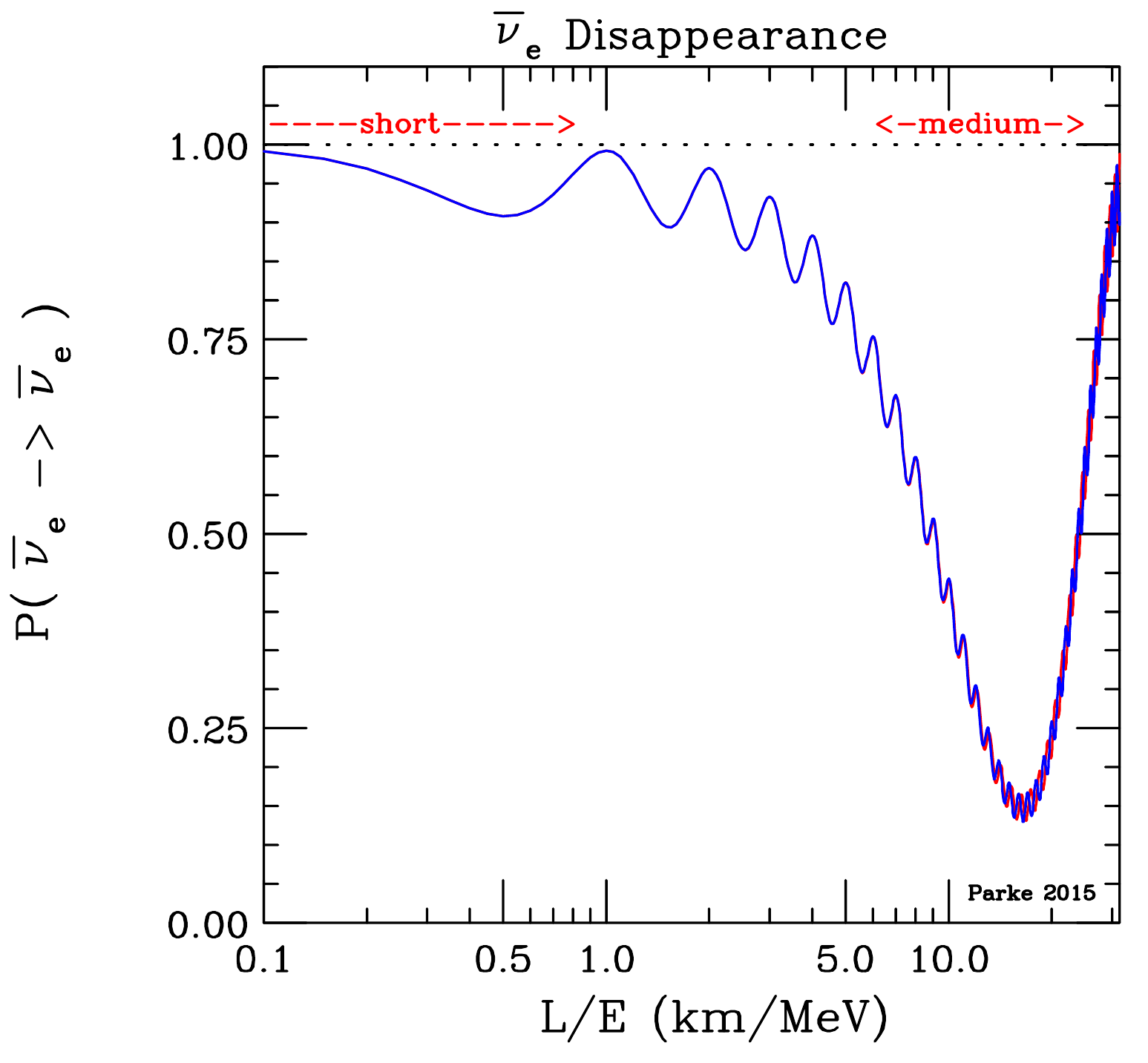}
\end{center}
\vspace{-0.5cm}
\caption{The vacuum survival probability for $\bar{\nu}_e$ as a function of $L/E$.  Blue is for the normal mass ordering (NO) and red is the inverted mass ordering (IO)
with  $\Delta m^2_{31}$  and $\Delta m^2_{32}$ chosen in such a fashion that the two survival probabilities are identical at small $L/E$, ie.
$\Delta m^2_{31} (IO) = - \Delta m^2_{31} (NO)+2 \sin^2 \theta_{12} \Delta m^2_{21}$.
Near the solar oscillation minimum, $L/E \sim 15$ km/MeV, the phase of the $\theta_{13}$ oscillations advances (retards) for the normal (inverted) mass ordering and the two oscillation probabilities are distinguishable, in principle. Also near the solar minimum, the amplitude of the $\theta_{13}$ oscillations is significantly reduced compared to smaller values of L/E.  The short baseline experiments, Daya Bay, RENO and Double Chooz, probe $L/E < 0.8$ km/MeV and the medium baseline, JUNO and RENO 50, probe $6 < L/E < 25 $ km/MeV, as indicated.}
\label{fig:prob}
\end{figure}

  \section{$\bar{\nu}_e$ Survival Probability in Vacuum:} 
\label{sec:npz}

  The exact $\bar{\nu}_e$ survival probability in vacuum, see Fig. \ref{fig:prob}, is given by\footnote{The standard PDG conventions with the kinematical phase given by $\Delta_{ij}\equiv \Delta m^2_{ij} L/4E$ or  $1.267 ~\Delta m^2_{ij} L/E$ depending on whether one is using natural or (eV$^2$, km, MeV) units. Also, matter effects shift the $\Delta m^2$ by $(1+{\cal O}(E/10 GeV))$, where $E<10$ MeV, so are negligible for typical reactor neutrinos experiments.} 
   \begin{eqnarray}
P_x(\bar{\nu}_e \rightarrow \bar{\nu}_e) & = &1- 4|U_{e2}|^2 |U_{e1}|^2 \sin^2 \Delta_{21} \nonumber \\[1mm]
& & -4|U_{e3}|^2 |U_{e1}|^2 \sin^2 \Delta_{31} -4|U_{e3}|^2 |U_{e2}|^2 \sin^2 \Delta_{32}  \nonumber \\[5mm]
& =& 1-  \cos^4\theta_{13} \sin^2 2 \theta_{12} \sin^2 \Delta_{21}     \nonumber \\[1mm]
 & & - ~\sin^2 2 \theta_{13}~ { ( \cos^2 \theta_{12} \sin^2 \Delta_{31}+\sin^2 \theta_{12} \sin^2 \Delta_{32}) },
 \label{eqn:P0a}
\end{eqnarray}
using $\Delta_{ij}\equiv \Delta m^2_{ij} L/4E$.

It was shown in \cite{Nunokawa:2005nx}, that to an excellent accuracy 
\begin{eqnarray}
 \cos^2 \theta_{12} \sin^2 \Delta_{31}+\sin^2 \theta_{12} \sin^2 \Delta_{32}  & \approx & \sin^2 \Delta_{ee} \nonumber \\
 {\rm where} ~ \Delta m^2_{ee}  & \equiv & \cos^2 \theta_{12}\Delta m^2_{31}+\sin^2 \theta_{12} \Delta m^2_{32}   \nonumber 
 \end{eqnarray}
for $L/E < 0.8$ km/MeV. A variant of this derivation is given in the Appendix \ref{A}.  

However, in this article we will use an exact formulation given in \cite{Minakata:2007tn} which follows Helmholtz, \cite{Helmholtz}, in combining the two oscillation frequencies, proportional to $\Delta_{31}$ and $\Delta_{32}$ into one frequency plus a phase.  The exact survival probability is given by (see Appendix \ref{B})
  \begin{eqnarray}
P_x(\bar{\nu}_e \rightarrow \bar{\nu}_e)  &= & 1- \cos^4\theta_{13} \sin^2 2 \theta_{12} \sin^2 \Delta_{21}     \nonumber \\[1mm]
 &   & - \frac{1}{2} \sin^2 2 \theta_{13} \biggl( 1 - \sqrt{ 1- \sin^2 2 \theta_{12} \sin^2 \Delta_{21}  } ~ \cos \Omega \biggr) 
 \label{eqn:P0}   \\[2mm]
{\rm with} \quad  \Omega & = & (\Delta_{31} +\Delta_{32}) + \arctan(\cos2\theta_{12} \tan \Delta_{21}).  \nonumber
 \end{eqnarray}
$\Omega$ consists of two parts: one that is even under the interchange of  $\Delta_{31}$ and $\Delta_{32}$ and is linear in $L/E$, $(\Delta_{31}+\Delta_{32})$, and the other which is odd under this interchange and contains both linear and higher (odd) powers in $L/E$, 
$\arctan(\cos2\theta_{12} \tan \Delta_{21})$, remember $\Delta_{21}=\Delta_{31}-\Delta_{32}$.

The key point is the separation of the kinematic phase, $\Omega$, into an effective $2\Delta$ (linear in $L/E$) and a phase, $\phi$. For short baseline experiments, it is natural to expand $\Omega$ in a power series in $L/E$ and identify the coefficient of the linear term in $L/2E$ as  the effective $\Delta m^2$ and include all the higher order terms in the phase\footnote{Appendix A of \cite{Minakata:2007tn} contains a discussion of an effective $\Delta m^2$ as a function of $L/E$ for arbitrary $L/E$. At $L/E=0$ this definition is identical to $\Delta m^2_{ee}$.  }. Then,
\begin{eqnarray} 
\Omega & = & 2 \Delta_{ee} + {\phi }  \label{eqn:omega} \\[3mm]
{\rm where}  \quad  \Delta m^2_{ee}  & \equiv  & \frac{\partial ~\Omega}{\partial (L/2E)} \left|_{\frac{L}{E} \rightarrow 0} \right.
 =  \cos^2 \theta_{12}\Delta m^2_{31}+\sin^2 \theta_{12} \Delta m^2_{32}  \label{eqn:dmsqee} 
\\[3mm]
{\rm and}  \quad \quad { \phi } & \equiv  & \Omega - 2 \Delta_{ee} 
=  \arctan(\cos 2 \theta_{12} \tan \Delta_{21}) - \Delta_{21} \cos 2 \theta_{12} .
\label{eqn:phi} 
\end{eqnarray}
With this separation, $2 \Delta_{ee}$ varies at the atmospheric scale, 0.5 km/MeV, whereas $\phi$ varies at the solar oscillation scale, 15 km/MeV, and 
\begin{eqnarray}
\phi =0, \quad \frac{\partial ~\phi}{\partial (L/2E)} =  0 \quad {\rm and} \quad   \frac{\partial^2 ~\phi}{\partial (L/2E)^2} =  0 \quad {\rm at} ~\frac{L}{E}=0,
\nonumber
 \end{eqnarray} 
therefore,  in a power series in $L/E$, $\phi$ starts at $(\Delta m^2_{21} L/E)^3$ (see eqn. \ref{eqn:phicubic}).

Since $\Omega$ only appears as $\cos \Omega$, it is useful to redefine $\Omega = 2 |\Delta_{ee}| \pm \phi$ , so that the sign associated with the mass ordering appears only in front of $\phi$.  If and only if this sign is determined, is the mass ordering determined in $\nu_e$ disappearance experiments.

\begin{figure}
\begin{center}
\vspace{-1cm}
\includegraphics[width=0.6\textwidth]{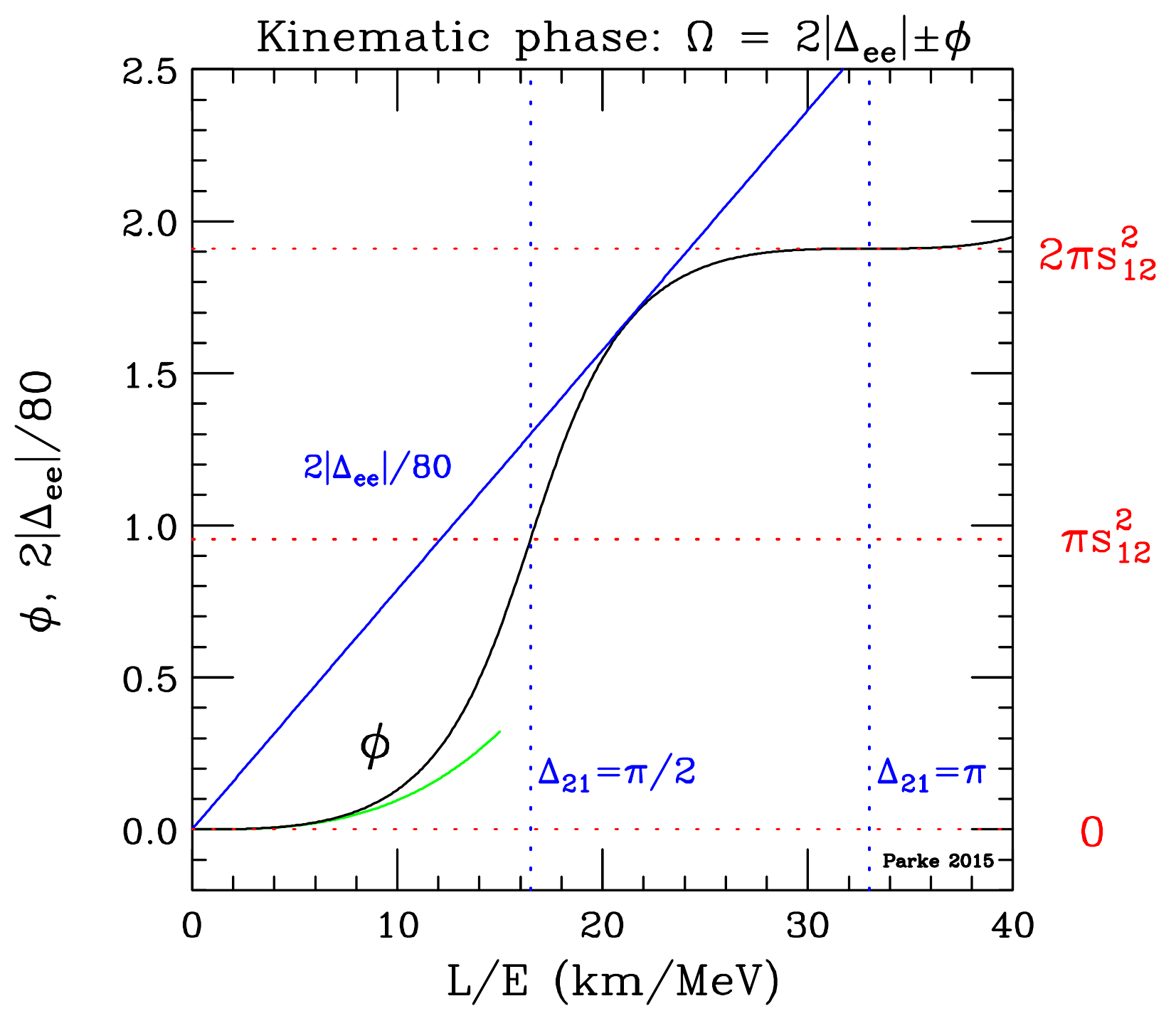}
\end{center}
\vspace{-0.5cm}
\caption{The $L/E$ dependence of the two components that make up the kinematic phase $\Omega=2|\Delta_{ee}| \pm \phi$ associated with the $\theta_{13}$ oscillation, eqn. \ref{eqn:P0}. $\phi$ is the black staircase function which increases by $2\pi \sin^2 \theta_{12}$ for every increase in $\Delta_{21}$ by $\pi$, see eqn. \ref{eqn:phi+pi}.
The blue straight line is $2|\Delta_{ee}|/80$, which is always greater than or equal to $\phi$.
The green curve is the $\Delta^3_{21}$ approximation to $\phi$ given in eqn. \ref{eqn:phicubic}, which is an excellent approximation for $L/E < 8$ km/MeV.
}
\label{fig:phi}
\end{figure}

There are three things worth noting about writing the exact $\nu_e$ survival probability as in eqn. \ref{eqn:P0}, with $\Omega$ given by eqn. \ref{eqn:omega}:
\begin{itemize}
\item The effective atmospheric $\Delta m^2$ associated with $\theta_{13}$ oscillation is a simple combination of the fundamental parameters,
 see eqn.  \ref{eqn:dmsqee} above or in ref. \cite{Nunokawa:2005nx} as they are identical,
\begin{eqnarray}
 \Delta m^2_{ee} & = & \cos^2 \theta_{12}\Delta m^2_{31}+\sin^2 \theta_{12} \Delta m^2_{32}  \nonumber  \\
  &=& \Delta m^2_{31} - \sin^2 \theta_{12} \Delta m^2_{21} =  \Delta m^2_{32} + \cos^2 \theta_{12} \Delta m^2_{21} \nonumber \\
  & = & m^2_3 - (\cos^2 \theta_{12} m^2_1 +\sin^2 \theta_{12} m^2_2). \nonumber
  \end{eqnarray}
Thus $ \Delta m^2_{ee} $ is simple the ``$\nu_e$ average of $ \Delta m^2_{31} $ and $ \Delta m^2_{32} $,'' since the $\nu_e$ ratio of $\nu_1$ to $\nu_2$ is  $\cos^2 \theta_{12}$ to  $\sin^2 \theta_{12}$, and determines the $L/E$ scale associated with the $\theta_{13}$ oscillations.  
\item
The modulation of the  amplitude associated with the $\theta_{13}$ oscillation,  is manifest in the square root multiplying the $\cos \Omega$ oscillating term, where
\begin{eqnarray}
 \sqrt{1- \sin^2 2 \theta_{12} \sin^2 \Delta_{21} } & =  & \left\{  
 \begin{array}{ll}
 1 & {\rm at} \quad \Delta_{21}=   n \pi \\
 \cos 2 \theta_{12} \approx 0.4 \quad   & {\rm at } \quad  \Delta_{21}= (2n+1) \pi/2 
 \end{array}
 \right.
 \label{eqn:ampmod}
\end{eqnarray}
for $n=0,1,2, \cdots$.  Thus, at solar oscillation minima, when $\Delta_{21}=0, \pi, 2\pi,...$, the oscillation amplitude is just $\sin^2 2\theta_{13}$, whereas at solar oscillation maxima,  when $\Delta_{21}=\pi/2, 3\pi/2,...$, the oscillation amplitude is $\cos 2\theta_{12} \sin^2 2\theta_{13}$ i.e. reduced by approximately 60\%.
\item
The phase, $\phi$, causes an advancement (retardation) of the $\theta_{13}$ oscillation for the normal (inverted) mass ordering of the
neutrino mass eigenstates. 
$\phi$  is a ``rounded'' staircase function\footnote{In the limit, $\sin^2 \theta_{12} \rightarrow \frac{1}{2}$, one recovers the well known result that this rounded staircase function becomes a true staircase or step function.}, which is zero and has zero first and second derivatives at $L/E=0$ ($\Delta_{21}=0$), but then between $L/E \sim 10-20$ km/MeV ($\Delta_{21} \sim \frac{\pi}{3} - \frac{2\pi}{3}$) rapidly jumps by $2 \pi \sin^2 \theta_{12}$, and this pattern is repeated for every increase of $L/E \sim 30$ km/MeV ($\Delta_{21}$ by $\pi$), i.e.
\begin{eqnarray}
\phi(\Delta_{21}\pm \pi) = \phi(\Delta_{21}) \pm 2 \pi \sin^2 \theta_{12},
\label{eqn:phi+pi}
\end{eqnarray}
see Fig. \ref{fig:phi}. Also shown on the same plot is $2|\Delta_{ee}|$ divided by 80. This number 80 was chosen so that $2|\Delta_{ee}|$ fits on the same plot and to demonstrate that $ 2|\Delta_{ee}| \ge 80 ~ \phi$ so that the shift in phase caused by $\phi$ is never bigger than a 1.25\% effect.  Also for $L/E < 5$ km/MeV, the shift in phase is much smaller than this, see next section. 
\end{itemize}


\subsection{Short Baseline Experiments ($0 < L/E < 1 $ km/MeV)}
For reactor experiments with baselines less than 2 km, the exact expression eqn. \ref{eqn:P0} contains 
elements which require measurement uncertainties on the oscillation probability to better than one part in $10^4$. This is way beyond the capability of the current or envisaged experiments. This occurs because for experiments at these baselines, the following conditions on the kinematic phases are satisfied,
\begin{eqnarray}
0< |\Delta_{31}| \approx |\Delta_{32}| < \pi  \quad \Rightarrow  0 < \Delta_{21} < 0.1,
\end{eqnarray}
and some elements of eqn. \ref{eqn:P0} dependent on higher powers of $\Delta_{21}$.
These elements are 
\begin{itemize}
\item The modulation of the $\theta_{13}$ oscillation amplitude which when expanded in powers of $\Delta_{21}$ is given by
\begin{eqnarray}
\sqrt{( 1- \sin^2 2 \theta_{12} \sin^2 \Delta_{21})  } & = & 1 -2 \sin^2 \theta_{12} \cos^2 \theta_{12} \Delta^2_{21} +{\cal O}(\Delta^4_{21}) \\
& = & 1+ {\cal O}(<10^{-3})
\end{eqnarray}
Remember, this amplitude modulation factor is multiplied by $\frac{1}{2}\sin^2 2\theta_{13} \sim 0.05$.  Reducing the effect of the amplitude modulation to less than one part in $10^4$.  
 
\item  The advancement or retardation of the kinematic phase, $\Omega$, caused by $\phi$ whose sign depends on the mass ordering.  For small values of $\Delta_{21}$ the advancing/retarding phase can be written as
\begin{eqnarray}
\phi &=&  \frac{1}{3}  \cos 2\theta_{12} \sin^2 2 \theta_{12} \Delta^3_{21} + {\cal O}(\Delta^5_{21}) 
\label{eqn:phicubic}
\end{eqnarray}
then using this approximation in the kinematic phase $\Omega$, we have
\begin{eqnarray}
\cos(2|\Delta_{ee}| \pm \phi ) &= & \cos( 2|\Delta_{ee}|) \cos \phi \mp \sin( 2|\Delta_{ee}|) \sin \phi  \nonumber \\
 & = & \cos( 2|\Delta_{ee}|)  \mp   \frac{1}{3}  \cos 2\theta_{12} \sin^2 2 \theta_{12} \Delta^3_{21} \sin( 2|\Delta_{ee}|)
 +  {\cal O}(\Delta^5_{21}) \nonumber \\ \label{eqn:omegacubic}   \\
& = & \cos( 2|\Delta_{ee}|) +{\cal O}(<10^{-4})  \nonumber 
\end{eqnarray}
Again remember, that we have a further reduction by $\frac{1}{2}\sin^2 2\theta_{13} \sim 0.05$.  Making the phase advancement or retardation significantly smaller than even the amplitude modulation for these experiments.
\end{itemize}

\begin{figure}
\begin{center}
\vspace{-1cm}
\includegraphics[width=0.6\textwidth]{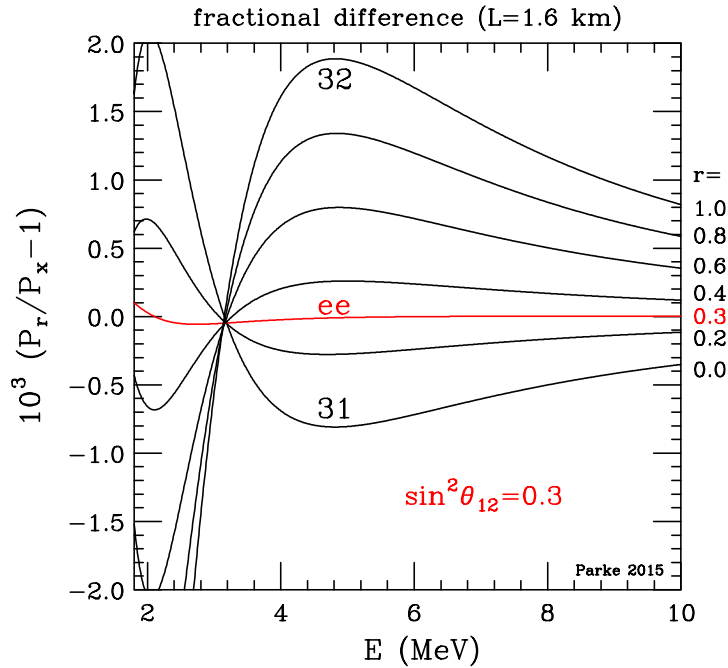}
\end{center}
\vspace{-0.5cm}
\caption{The fractional difference between the exact survival probability, eqn. \ref{eqn:P0a}, and a sequence of approximate survival probabilities, 
where 
$\cos^2 \theta_{12} \sin^2 \Delta_{31} +\sin^2 \theta_{12} \sin^2 \Delta_{32}$ is replaced with $\sin^2 ( \Delta m^2_{rr}L/4E) $ with $\Delta m^2_{rr} \equiv (1-r)\Delta m^2_{31}+ r \Delta m^2_{32}$.  Clearly, $r=\sin^2 \theta_{12}$ minimizes the absolute value of the fractional difference between the exact and approximate survival probabilities.  Thus, the approximation  of replacing  $\cos^2 \theta_{12} \sin^2 \Delta_{31} +\sin^2 \theta_{12} \sin^2 \Delta_{32}$ with $\sin^2 \Delta_{ee}$ gives an approximate survival probability  that is better than one part in $10^{4}$ over the L/E range of the Daya Bay, RENO and Double Chooz experiments.
  }
\label{fig:Pshort}
\end{figure}

Using this information in the $\nu_e$ survival probability, we can replace eqn. \ref{eqn:P0} by
\begin{eqnarray}
P_{\rm short}(\bar{\nu}_e \rightarrow \bar{\nu}_e) &= & 1- \cos^4\theta_{13} \sin^2 2 \theta_{12} \sin^2 \Delta_{21}    
 - \sin^2 2 \theta_{13}  \sin^2 | \Delta_{ee}| . 
 \label{eqn:Pshort}  
 \end{eqnarray}
 which is accurate to better than one part in $10^{4}$. 
In Fig. \ref{fig:Pshort} the fractional difference between eqn. \ref{eqn:P0} and \ref{eqn:Pshort} is shown for an experiment with a baseline of 1.6 km.  Since the measurement uncertainty on the $\nu_e$ survival probability is much greater  ($>$~0.01\%)  than the difference between the exact, eqn. \ref{eqn:P0}, and the approximate, eqn. \ref{eqn:Pshort},  survival probabilities, use of either will result in the same measured values of the parameters 
$\sin^2 2 \theta_{13}$ and $|\Delta m^2_{ee}|$ i.e. the measurement uncertainties will dominate. 

If new, extremely precise, short baseline experiments ever need a more accurate survival probability, one could easily add the first correction of the amplitude modulation, giving
\begin{eqnarray}
P_{\rm xshort}(\bar{\nu}_e \rightarrow \bar{\nu}_e) &= & 1- \cos^4\theta_{13} \sin^2 2 \theta_{12} \sin^2 \Delta_{21}    
 \nonumber \\[1mm] &   & 
\hspace*{-1cm}
 - \sin^2 2 \theta_{13} ~[~ \sin^2 |\Delta_{ee}| 
 + \sin^2 \theta_{12} \cos^2 \theta_{12} \Delta^2_{21}  \cos(2|\Delta_{ee}|)~] 
 \label{eqn:Pshort2}  
 \end{eqnarray}
and this would improve the accuracy of the approximation to better than one part in $10^{5}$. 

An alternative way to derive these approximate survival probability, eqn \ref{eqn:Pshort} \& \ref{eqn:Pshort2}, is given in the Appendix \ref{A}.

\section{Other Possible Definitions of an Effective $\Delta m^2$}
\label{sec:otherdmsqs}

\subsection{A New Definition of the Effective $\Delta m^2$}

Another possible way to define an effective $\Delta m^2$, here I will use the symbol $\Delta m^2_{XX}$, is as follows
\begin{eqnarray}
\Delta m^2_{XX} & \equiv & \sqrt{ \cos^2 \theta_{12} ~(\Delta m^2_{31})^2+  \sin^2 \theta_{12} ~(\Delta m^2_{32})^2~ }.
\end{eqnarray}
Clearly this definition is independent of $L/E$ and it guarantees that, in the limit $L/E \rightarrow 0$, that
\begin{eqnarray}
\sin^2 \Delta_{XX} & =& \cos^2 \theta_{12} ~\sin^2 \Delta_{31}+  \sin^2 \theta_{12} ~\sin^2 \Delta_{32}.
\label{eqn:XXdefn}
\end{eqnarray}
One can then show, that
\begin{eqnarray}
|\Delta m^2_{XX}|  & =  & |\Delta m^2_{ee}|~(1+ {\cal O}\biggr[~ \left( \frac{\Delta m^2_{21}} {\Delta m^2_{ee}} \right)^2 ~\biggr]) \end{eqnarray}
So $|\Delta m^2_{XX}|$ is essentially equal to  $|\Delta m^2_{ee}|$ up to correction on the order of $10^{4}$, including the effects of the solar mixing angle\footnote{The following, useful identity is easy to prove by writing $\Delta m^2_{21}= \Delta m^2_{31}- \Delta m^2_{32}$:\\ 
$(\cos^2\theta_{12} \Delta m^2_{31}+ \sin^2\theta_{12} \Delta m^2_{32})^2 =
[\cos^2\theta_{12} (\Delta m^2_{31})^2+ \sin^2\theta_{12} (\Delta m^2_{32})^2] - \cos^2 \theta_{12} \sin^2 \theta_{12} (\Delta m^2_{21})^2$.}.

A variant of this definition of an effective $\Delta m^2$ (here I will used the subscripts ``$xx$''), is defined in terms of the position of the first extremum of $ ( \cos^2 \theta_{12} ~\sin^2 \Delta_{31}+  \sin^2 \theta_{12} ~\sin^2 \Delta_{32} )$ in $L/E$.  If this extremum occurs at $(L/E)|_1$, then define
\begin{eqnarray}
\Delta m^2_{xx} & \equiv & {2\pi \over (L/E)|_1},
\end{eqnarray}
so that, at this extremum, $ \frac{\Delta m^2_{xx} L}{4 E}  =  \frac{\pi}{2}$.   With this definition it is again easy to show that, 
\begin{eqnarray}
|\Delta m^2_{xx}|  & =  & |\Delta m^2_{ee}|~(1+ {\cal O}\biggr[~ \left( \frac{\Delta m^2_{21}} {\Delta m^2_{ee}} \right)^2 ~\biggr]). \end{eqnarray}
Again, essentially equal to  $\Delta m^2_{ee}$.

In both $|\Delta m^2_{XX}|$  and $|\Delta m^2_{xx}|$, the corrections of order $\left( \frac{\Delta m^2_{21}} {\Delta m^2_{ee}} \right)^2$, come from the amplitude modulation of the $\theta_{13}$ oscillation and the coefficients are $ \frac{1}{2} \sin^2 \theta_{12} \cos^2 \theta_{12}$ and $ \sin^2 \theta_{12} \cos^2 \theta_{12}$ respectively.  Note, these corrections are mass ordering independent.

\subsection{Daya Bay's Original Definition of the Effective $\Delta m^2$}
In ref. \cite{An:2013zwz} \& \cite{An:2015rpe}, the Daya Bay experiment used the following definition for an effective $\Delta m^2$, here I will use the symbol  $\Delta m^2_{YY}$,
\begin{eqnarray}
\sin^2 \Delta_{YY} & \equiv & \cos^2 \theta_{12} ~\sin^2 \Delta_{31}+  \sin^2 \theta_{12} ~\sin^2 \Delta_{32}.
\label{eqn:dayabayee}
\end{eqnarray}
which implies that
 \begin{eqnarray}
\Delta m^2_{YY} & \equiv & \left( \frac{4E}{L}\right)  \arcsin \left[ \sqrt{(\cos^2 \theta_{12} \sin^2 \Delta_{31}+ \sin^2\theta_{12} \sin^2 \Delta_{32}) }
\right].
\label{eqn:dayabayee1}
\end{eqnarray} 
For $L/E < 0.3$ km/MeV, so that $\sin^2 \Delta_{3i} = \Delta^2_{3i}$ is a good approximation, $\Delta m^2_{YY}$
is approximately independent of $L/E$. However, for larger values of L/E, $\Delta m^2_{YY}$ is L/E dependent, exactly in the L/E region, $0.3 < L/E < 0.7$ km/MeV, where the  bulk of the experimental data from the far detectors of the Daya Bay experiment is obtained.  In the center of this L/E region, $L/E \approx 0.5$ km/MeV, is the position of the oscillation minimum.

\begin{figure}
\begin{center}
\vspace{-1cm}
\includegraphics[width=0.6\textwidth]{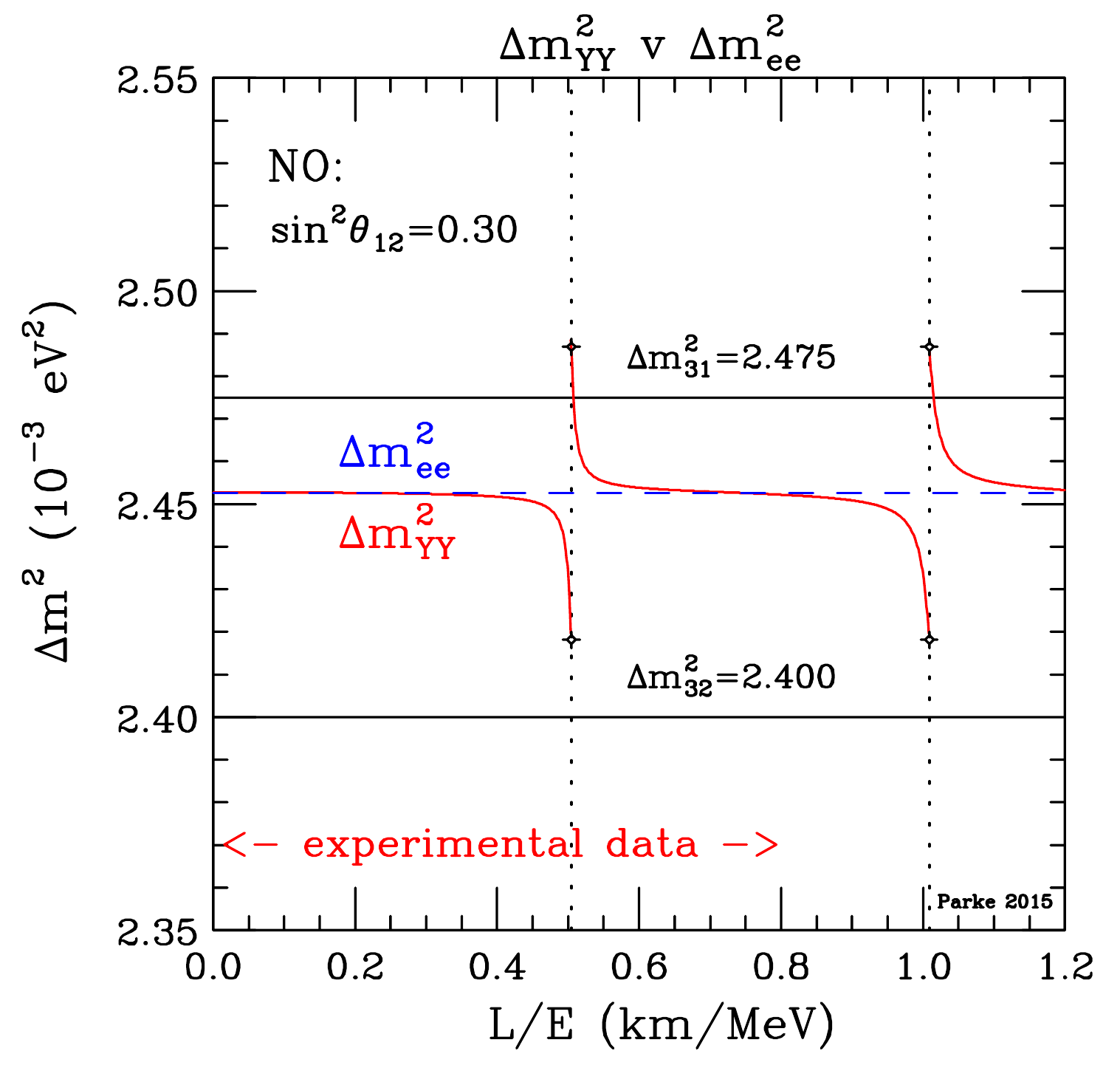}
\end{center}
\vspace{-1cm}
\caption{Daya Bay's original definition, see \cite{An:2013zwz} and \cite{An:2015rpe}, for an effective $\Delta m^2$,  $\Delta m^2_{YY}$, is given by the solid red line. Notice the sizeable L/E dependence near oscillation minimum and maximum (vertical black dotted lines). At all oscillation extrema, this definition is discontinuous and the size of the discontinuity is  $~\sin 2 \theta_{12} \Delta m^2_{21}  \sim 3\%$. The first discontinuity occurs in the middle of the experimental data of the Daya Bay, RENO and Double Chooz experiments. The L/E independent lines: $\Delta m^2_{ee} \equiv \cos^2 \theta_{12}\Delta m^2_{31}+\sin^2 \theta_{12} \Delta m^2_{32}$ is the blue dashed, $\Delta m^2_{31}$ and $\Delta m^2_{32}$ are the labelled black lines. This figure is for normal mass ordering with $\sin^2 \theta_{12}=0.30$ and  $\Delta m^2_{ee}= 2.453 \times 10^{-3}$ eV$^2$.
}
\label{fig:dmsq}
\end{figure}

Furthermore, the definition given by Eqn. \ref{eqn:dayabayee}, is discontinuous at oscillation minimum (OM).  This occurs because as you increase $L/E$, the L.H.S.  eqn. \ref{eqn:dayabayee} can go to 1, whereas the R.H.S. never reaches 1.  So to satisfy Eqn. \ref{eqn:dayabayee}, as you increase $L/E$, your effective $\Delta m^2$ must be discontinuous at OM and the size of this discontinuity is given by\footnote{The following identity is useful to understand this point, $\sin^2(\frac{\pi}{2} \pm \epsilon)\approx  1-\epsilon^2$ where here $\epsilon=s_{12} c_{12} \Delta_{21}$. Similarly at oscillation maximum, $\sin^2(\pi \pm \epsilon) \approx  \epsilon^2$. }
\begin{eqnarray}
\delta  ~\Delta m^2_{EE}|_{OM} & = &  \sin 2\theta_{12}  \Delta m^2_{21}
\end{eqnarray}
which is of order of 3\%.  In Fig. \ref{fig:dmsq}, the various $\Delta m^2$'s are plotted as a function of L/E.

The relationship between Daya Bay's $\Delta m^2_{YY}$  and that of the previous section is as follows
\begin{eqnarray}
\Delta m^2_{YY}|_{L/E \rightarrow 0}  = \Delta m^2_{ee}\sqrt{\left( 1+  \sin^2\theta_{12} \cos^2 \theta_{12} \left(\frac{\Delta m^2_{21}}{\Delta m^2_{ee}}\right)^2 \right) }  ~.
\end{eqnarray}
Therefore they are identical up to corrections of ${\cal O}(10^{-4})$ as L/E $\rightarrow 0$.

Given that $\Delta m^2_{YY}$  is L/E dependent one should take the average of $\Delta m^2_{YY}$ over the L/E range of the experiment
\begin{eqnarray}
\langle \Delta m^2_{YY} \rangle  & = & 
\frac{ \int_{(L/E)_{min}}^{(L/E)_{max}} d(L/E)  ~\Delta m^2_{YY} }{ [(L/E)_{max} - (L/E)_{min}] } ~.
\end{eqnarray}
For the current experiments this range is from [0,0.8] km/MeV and then from Fig, \ref{fig:dmsq} it is clear that
\begin{eqnarray}
\langle \Delta m^2_{YY} \rangle  & \approx & \Delta m^2_{ee},
\end{eqnarray}
if the discontinuity at OM is averaged over in a symmetric way. In practice, of course, one needs to weight the average over the L/E range by the experimental L/E sensitivity.  This is something that can only be performed by the experiment.  This was not performed in ref. \cite{An:2013zwz} or \cite{An:2015rpe}.

\begin{figure}
\begin{center}
\vspace{-1cm}
\includegraphics[width=0.6\textwidth]{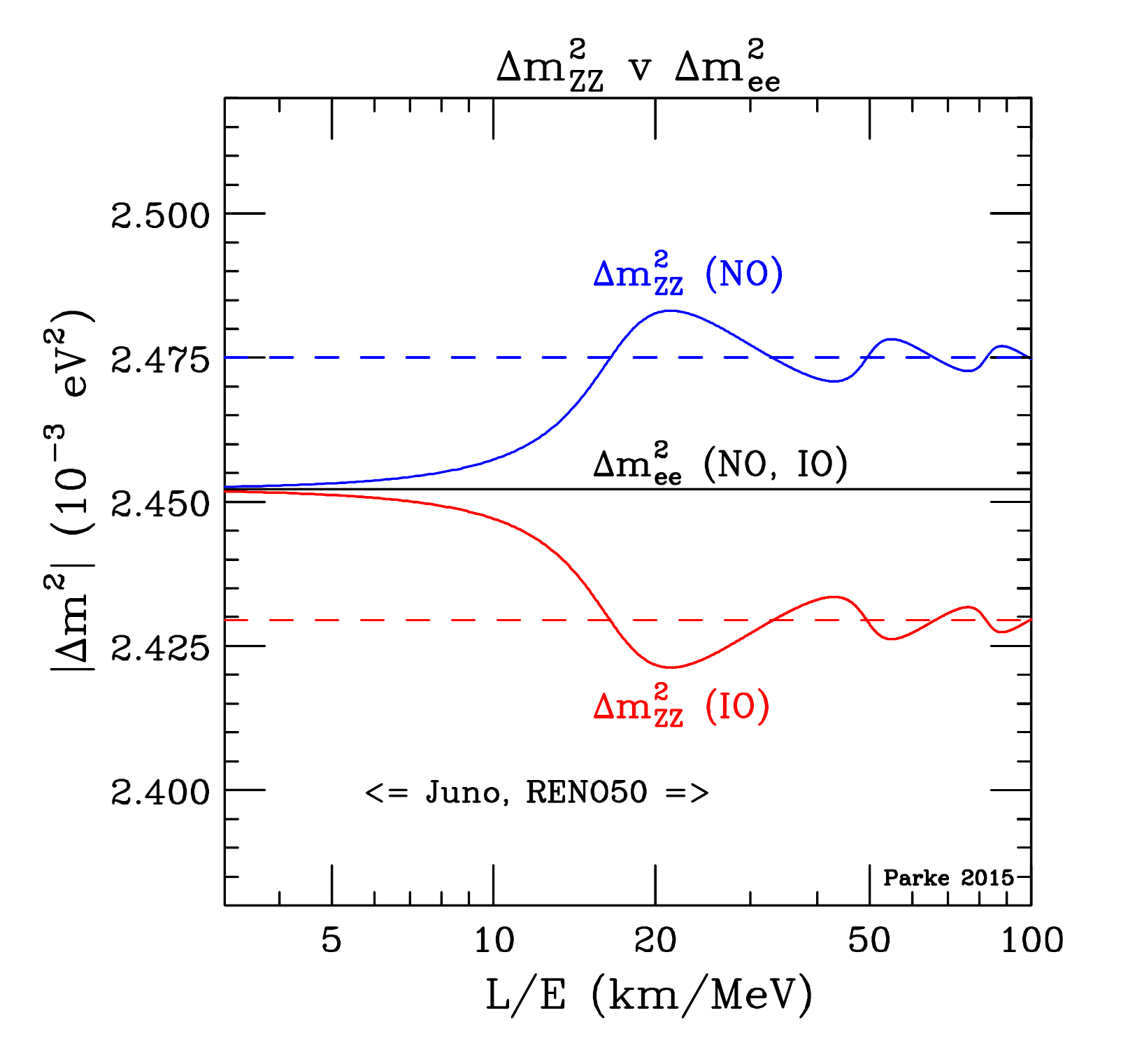}
\end{center}
\vspace{-0.5cm}
\caption{Daya Bay's new definition, see \cite{An:2015rpev2}, of an effective $\Delta m^2$,  $\Delta m^2_{ZZ}$, for $\bar{\nu}_e$ disappearance compared $\Delta m^2_{ee}\equiv \cos^2 \theta_{12}\Delta m^2_{31}+\sin^2 \theta_{12} \Delta m^2_{32}$. The L/E range appropriate for JUNO and RENO-50 is 6 to 25 km/MeV, exactly the range in which  $\Delta m^2_{ZZ}$ changes by $\pm$1\%. Yet, the expected accuracy of these two experiments is better than 0.5\%.  The sign of the variation of  $\Delta m^2_{ZZ}$ is mass ordering dependent. The blue and red dashed lines are $\Delta m^2_{31}$ for NO and IO respectively.}
\label{fig:JUNORENO50}
\end{figure}

\subsection{Daya Bay's New Definition of the Effective $\Delta m^2$}
After the issue with  $\Delta m^2_{YY}$ was pointed out to the Daya Bay collaboration \cite{pc20150606}, the Daya Bay collaboration defined a new effective $\Delta m^2$ in the supplemental material of ref. \cite{An:2015rpev2}.  Here I will use the symbol  $\Delta m^2_{ZZ}$ for this new definition which is defined in terms of the kinematic phase, $\Omega$, given eqn. \ref{eqn:omega}, as
\begin{eqnarray}
\Delta m^2_{ZZ} & \equiv & \frac{2E}{L} ~\Omega,  \\
& = & |\Delta m^2_{ee}| \pm \frac{2E}{L} ~\phi. \nonumber
\end{eqnarray}
Unfortunately, since $\phi$ is not a linear function in $L/E$, $\Delta m^2_{ZZ}$ is also $L/E$ dependent.
In contrast remember, from eqn. \ref{eqn:dmsqee}, $ \Delta m^2_{ee}   \equiv   \frac{\partial ~\Omega}{\partial (L/2E)} \left|_{\frac{L}{E} \rightarrow 0} \right. $.

For short baseline experiments, such as Daya Bay, RENO and Double Chooz, this dependence is small, and can be calculated analytically from eqn. (\ref{eqn:phicubic}),
\begin{eqnarray} 
\Delta m^2_{ZZ}   & = & |\Delta m^2_{ee}| \left[1\pm  \frac{1}{6}  \cos 2\theta_{12} \sin^2 2 \theta_{12} \left( \frac{\Delta m^2_{21}}{\Delta m^2_{ee}} \right)   \Delta^2_{21} +{\cal O}\left( \left( \frac{\Delta m^2_{21}}{\Delta m^2_{ee}} \right)  \Delta^4_{21}\right)   \right]  \nonumber \\
 & \approx & |\Delta m^2_{ee}| \left[1\pm 6 \times 10^{-6} \left(\frac{L/E}{0.5~{\rm km/MeV}}\right)^2 \right] .
\end{eqnarray}
Given the current and expected future accuracy of the current short baseline experiments, the L/E dependence in $\Delta m^2_{ZZ}$ can be ignored.

However for future experiments such as JUNO,  \cite{An:2015jdp}, and RENO-50,  \cite{reno50}, the L/E dependence of $\Delta m^2_{ZZ}$ is significant, see Fig. \ref{fig:JUNORENO50}.  These experiments explore an L/E range from 6 to 25 km/MeV. In this range, $\Delta m^2_{ZZ}$ changes by $\sim 1$\% whereas the expected accuracy of the measurement is better than 0.5\%, see \cite{An:2015jdp}.  So this definition of  $\Delta m^2_{ZZ}$ is not appropriate for these experiments unless the experiments want to do the L/E averaging as discussed in the previous section.

\section{Conclusions}

Having a single, $L/E$ independent effective $\Delta m^2$ which can be used for reactor experiments of any $L/E$ is highly desirable.  $\Delta m^2_{ee}$, defined in eqn. \ref{eqn:dmsqee}, is  the best  effective  $\Delta m^2$ for $\nu_e$ disappearance in the literature for the following reasons:
\begin{itemize}
\item  Is independent of $L/E$ for all values of $L/E$.
\item Is a simple combination of fundamental parameters:
\begin{eqnarray}
\Delta m^2_{ee} & \equiv & \cos^2 \theta_{12} ~\Delta m^2_{31} + \sin^2 \theta_{12} ~\Delta m^2_{32} \nonumber \\
& =  & \Delta m^2_{31}-\sin^2 \theta_{12} ~\Delta m^2_{21}= \Delta m^2_{32}+\cos^2 \theta_{12} ~\Delta m^2_{21}.  \nonumber \\
& = & m^2_3-(\cos^2 \theta_{12} m^2_1 +\sin^2 \theta_{12} m^2_2). \nonumber
\end{eqnarray}
\item Has a direct, simple, physical interpretation:\\ $ \Delta m^2_{ee}$ is ``the $\nu_e$ weighted average of $\Delta m^2_{31}$ and $\Delta m^2_{32}$,''
 since the ratio of the $\nu_e$ fraction in $\nu_1:\nu_2$ is $\cos^2 \theta_{12}:\sin^2 \theta_{12}$.
\item Can be used in the future medium baseline reactor experiments, $L/E > 6$ and $< 25$ km/MeV, using the exact oscillation probability,
\begin{eqnarray}
P(\bar{\nu}_e \rightarrow \bar{\nu}_e) & = & 1-  \cos^4 \theta_{13}\sin^2 2\theta_{12}  \sin^2 \Delta_{21} \nonumber \\
& & \hspace{-1cm} -\frac{1}{2} \sin^2 2 \theta_{13}\biggr(1-\sqrt{1-\sin^2 2 \theta_{12} \sin^2 \Delta_{21}}  ~\cos(2|\Delta_{ee}|\pm\phi) \biggr),  
\nonumber
\end{eqnarray}
where $ \phi  \equiv   \arctan( \cos 2\theta_{12} \tan \Delta_{21}) - \Delta_{21} \cos 2\theta_{12}$.
This probability can be used to determine  solar parameters $\sin^2 \theta_{12}$ and $\Delta m^2_{21}$ as well as $|\Delta m^2_{ee}|$ with unprecedented precision and may be able to determine the atmospheric mass ordering, if the sign in front of $\phi$ can be determined at high enough confidence level.
\item
Can be used in the current short baseline reactor experiments, $L/E < 1$ km/MeV, using the approximate oscillation probability,
\begin{eqnarray}
P(\bar{\nu}_e \rightarrow \bar{\nu}_e) & \approx & 1- \cos^4 \theta_{13}\sin^2 2\theta_{12} \sin^2 \Delta_{21}  - \sin^2 2\theta_{13}  \sin^2 \Delta_{ee}.
\nonumber
\end{eqnarray}
This is trivially obtained from the exact expression, eqn. \ref{eqn:P0}, by setting both the amplitude modulation to one and the phase advancement or retardation to zero, \\
\begin{eqnarray}
 \sqrt{1-\sin^2 2\theta_{12} \sin^2 \Delta_{21}} ~\rightarrow 1 \quad {\rm  and} \quad  \phi \rightarrow 0
 \nonumber
 \end{eqnarray}
 as these are higher order effects.
This approximates the exact oscillation probability to better than 1 part in $10^4$ and can be improved in a systematic way, see Eqn. \ref{eqn:taylor}. This probability, using the current experimental data, allows for an accurate determination of 
mixing angle $\theta_{13}$ and the atmospheric mass splitting $|\Delta m^2_{ee}|$,  independent of the atmospheric mass ordering, and only very weakly dependent on our current knowledge of the solar parameters, through the solar term.  
From a measured value of  $|\Delta m^2_{ee}|$, using short baseline reactor experiments,  it is simple to calculate  $\Delta m^2_{31}$ for both mass orderings. However, the uncertainties on $\Delta m^2_{31}$ will be more dependent on solar parameters, measured by other experiments, than $|\Delta m^2_{ee}|$.
\end{itemize}

Furthermore, $\Delta m^2_{ee}$, defined by eqn. \ref{eqn:dmsqee}, naturally appears as the renormalized atmospheric $\Delta m^2$ in neutrino propagation in matter, see \cite{Minakata:2015gra}, as using this renormalized $\Delta m^2$ significantly reduces the complexity of the oscillation probabilities.

\begin{acknowledgments}
The author acknowledges discussions with the Kam-Biu Luk, Yasuhiro Nakajima, Daniel Dwyer of the Daya Bay collaboration, Soo-Bong Kim and Seon-Hee Seo from the RENO collaboration as well as my collaborators Hisakazu Minakata and Peter Denton.
The author also acknowledges partial support from the  European Union FP7  ITN INVISIBLES (Marie Curie Actions, PITN- GA-2011- 289442). Fermilab is operated by the Fermi Research Alliance under contract no. DE-AC02-07CH11359 with the U.S. Department of Energy. 
\end{acknowledgments}


\section{Appendix A}
\label{A}
In this Appendix, an alternative derivation of why $\sin^2 \Delta_{ee}$ is the most accurate approximation for  $\cos^2 \theta_{12} \sin^2 \Delta_{31} + \sin^2 \theta_{12} \sin^2 \Delta_{32} $ is given.  Starting with the following linear combination of $\Delta m^2_{31}$ and  $\Delta m^2_{32}$, given by
\begin{eqnarray}
 \Delta_{rr} \equiv (1-r)\Delta_{31} + r\Delta_{32} \quad {\rm then} \quad  \Delta_{31} & = & \Delta_{rr}+r \Delta_{21}, \quad  \Delta_{32}=\Delta_{rr}-(1-r)\Delta_{21},   \nonumber 
\end{eqnarray}
since $\Delta_{21}=\Delta_{31}-\Delta_{32}$ and  r is a number between [0,1].  The relevant range of kinematic phases is $0\leq | \Delta_{31}| \sim |\Delta_{32}| <\pi $ and $0 \leq \Delta_{21} < \pi/30 \approx 0.1$.
So it is a simple exercise to perform a Taylor series expansion about $\Delta_{rr}$ using expansion parameter $\Delta_{21}$, and obtain  (using $c^2_{12} \equiv \cos^2 \theta_{12}$ and $s^2_{12} \equiv \sin^2 \theta_{12}$)
\begin{eqnarray}
c^2_{12} \sin^2 \Delta_{31} + s^2_{12} \sin^2 \Delta_{32} & = &   \sin^2 \Delta_{rr}  \nonumber \\  
& & + [c^2_{12} r -s^2_{12} (1-r) ] ~\Delta_{21}  \sin (2 \Delta_{rr})  \nonumber \\  
& &+  [c^2_{12} r^2 + s^2_{12} (1-r)^2] ~\Delta^2_{21} \cos(2\Delta_{rr})  \nonumber \\
& & -\frac{2}{3}[ c^2_{12} r^3 - s^2_{21} (1-r)^3]~\Delta^3_{21} \sin(2\Delta_{rr}) \nonumber \\
& & -\frac{1}{3} [c^2_{12} r^4 +s^2_{12} (1-r)^4] ~\Delta^4_{21} \cos(2\Delta_{rr}) \nonumber \\
& & + {\cal O}(\Delta^5_{21}). 
\end{eqnarray}
The choice of $r=s^2_{12}$, making $\Delta_{rr} = \Delta_{ee}$, does two great things for this Taylor series expansion:
\begin{enumerate}
\item the coefficient of $\Delta_{21}$ vanishes, since $[c^2_{12} r -s^2_{12} (1-r)]=0$,
\item and, the coefficient of $\Delta^2_{21}$ is a minimized, since 
\begin{eqnarray}
\left. \frac{\partial}{\partial r} [c^2_{12} r^2 + s^2_{12} (1-r)^2] \right|_{r=s^2_{12}} =0 \quad {\rm and} \quad  \frac{\partial^2}{\partial^2 r} [c^2_{12} r^2 + s^2_{12} (1-r)^2] > 0.  \nonumber
 \end{eqnarray}
\end{enumerate}
No other value of $r$ satisfies either of these requirements.  Thus, using $r=s^2_{12}$ makes $\sin^2 \Delta_{ee}$ the best possible approximation to $c^2_{12} \sin^2 \Delta_{31} + s^2_{12} \sin^2 \Delta_{32}$ for a constant $\Delta m^2$ and the corrections are tiny, of ${\cal O}(10^{-3})$ for L/E $<$ 1 km/MeV.

Using this expansion the $\nu_e$ survival probability can be written as
\begin{eqnarray}
P_{\rm xshort}(\bar{\nu}_e \rightarrow \bar{\nu}_e) &= & 1- \cos^4\theta_{13} \sin^2 2 \theta_{12} \sin^2 \Delta_{21}    
 \nonumber \\[1mm] &   & 
 - \sin^2 2 \theta_{13} ~\biggr[~ \sin^2 |\Delta_{ee}|   \nonumber  \\
& & \quad \quad  + \sin^2 \theta_{12} \cos^2 \theta_{12} \Delta^2_{21}  \cos(2|\Delta_{ee}|)~] \nonumber \\
& &\quad \quad  \mp~\frac{1}{6} \cos 2\theta_{12} \sin^2 2\theta_{12}~\Delta^3_{21} \sin(2|\Delta_{ee}|) \nonumber  \\
& & \quad \quad - ~\frac{1}{48}  \sin^2 2\theta_{12}~ [4 -3 \sin^2 2\theta_{12}] ~\Delta^4_{21} \cos(2|\Delta_{ee}|) \nonumber \\
& & \quad \quad + ~{\cal O}(\Delta^5_{21})~ \biggr]. \label{eqn:taylor}
\end{eqnarray}

\section{Appendix B}
\label{B}
The simplist way to show that
\begin{eqnarray}
\cos^2\theta_{12} \sin^2 \Delta_{31}+\sin^2\theta_{12} \sin^2 \Delta_{32} & = & \frac{1}{2}  \biggl( 1 - \sqrt{ 1- \sin^2 2 \theta_{12} \sin^2 \Delta_{21}  } ~ \cos \Omega \biggr) 
\end{eqnarray}
 with 
\begin{eqnarray} 
\Omega & = & 2 \Delta_{ee} + {\phi }  \\[3mm]
{\rm where}  \quad  \Delta m^2_{ee}  & \equiv  & \frac{\partial ~\Omega}{\partial (L/2E)} \left|_{\frac{L}{E} \rightarrow 0} \right.
 =  \cos^2 \theta_{12}\Delta m^2_{31}+\sin^2 \theta_{12} \Delta m^2_{32} 
\\[3mm]
{\rm and}  \quad \quad { \phi } & \equiv  & \Omega - 2 \Delta_{ee} 
= \arctan(\cos 2 \theta_{12} \tan \Delta_{21}) - \Delta_{21} \cos 2 \theta_{12},
\end{eqnarray}
is to write  
\begin{eqnarray}
c^2_{12} \sin^2 \Delta_{31}+s^2_{12} \sin^2 \Delta_{32} &= & \frac{1}{2}  \biggl( 1- (c^2_{12} \cos 2\Delta_{31}+s^2_{12} \cos 2\Delta_{32}) \biggr),
\end{eqnarray}
using $c^2_{12} \equiv \cos^2 \theta_{12}$ and $s^2_{12} \equiv \sin^2 \theta_{12}$.

Then, if we rewrite $2\Delta_{31}$ and $2\Delta_{32}$ in terms of $(\Delta_{31}+\Delta_{32})$ and $\Delta_{21}$, 
we have
\begin{eqnarray}
c^2_{12} \cos 2\Delta_{31}+s^2_{12} \cos 2\Delta_{32} & = & 
c^2_{12} \cos (\Delta_{31}+\Delta_{32}+\Delta_{21})+s^2_{12} \cos (\Delta_{31}+\Delta_{32}-\Delta_{21}) \nonumber \\
& = & \cos(\Delta_{31}+\Delta_{32}) \cos \Delta_{21} - \sin(\Delta_{31}+\Delta_{32}) \cos 2 \theta_{12} \sin \Delta_{21}.  \nonumber 
\end{eqnarray}
Since
\begin{eqnarray}
 \cos^2 \Delta_{21}+ \cos^2 2 \theta_{12} \sin^2 \Delta_{21} =1- \sin^2 2 \theta_{12} \sin^2 \Delta_{21} \nonumber
 \end{eqnarray}
 we can then write
\begin{eqnarray}
c^2_{12} \cos 2\Delta_{31}+s^2_{12} \cos 2\Delta_{32} 
& = & \sqrt{ 1- \sin^2 2 \theta_{12} \sin^2 \Delta_{21}  }  ~ \cos \Omega,  
\end{eqnarray}
where
\begin{eqnarray}
\Omega & = & \Delta_{31}+ \Delta_{32} + \arctan(\cos2\theta_{12} \tan \Delta_{21}). \nonumber
\end{eqnarray}
Applying the prescription given in Sec. \ref{sec:npz} to sepearate $\Omega$ into an effective $2\Delta$ and a phase, $\phi$, we find
\begin{eqnarray}
 \frac{\partial ~\Omega}{\partial L/2E} \left|_{\frac{L}{E} \rightarrow 0} \right.
& =  &   \cos^2 \theta_{12}\Delta m^2_{31}+\sin^2 \theta_{12} \Delta m^2_{32} =\Delta m^2_{ee}  \nonumber \\[3mm]
{\rm and} \quad \phi  & = &  \Omega - 2 \Delta_{ee}  
=  \arctan(\cos 2 \theta_{12} \tan \Delta_{21}) - \Delta_{21} \cos 2 \theta_{12}   \nonumber
\end{eqnarray}
thus
\begin{eqnarray}
\Omega & = & 2 \Delta_{ee} +  (\arctan(\cos2\theta_{12} \tan \Delta_{21})  -  \Delta_{21} \cos2\theta_{12} ),
\end{eqnarray}
qed.

\end{document}